\documentclass[a4, twocolumn, showpacs,prl ]{revtex4}
\usepackage{times}
\usepackage{graphics}
\usepackage{graphicx}
\usepackage{dcolumn}
\usepackage{amsmath}

\begin{document}
\bibliographystyle{apsrev}

\title{Frequency stabilization of an ultraviolet laser to ions in a discharge}
\author{E.W. Streed}
\email{ e.streed@griffith.edu.au}
\author{T.J. Weinhold}
\author{D. Kielpinski}

\affiliation{Centre for Quantum Dynamics, Griffith University, Nathan QLD 4111, Australia }
\date{\today}

\begin{abstract}
We stabilize an ultraviolet diode laser system at 369.5 nm to the optical absorption signal from  $\mbox{Yb}^+$ ions in a hollow-cathode discharge lamp. The error signal for stabilization is obtained by Zeeman spectroscopy of the 3 GHz-wide absorption feature. The frequency stability is independently measured by comparison against the fluorescence signal from a laser-cooled crystal of $^{174}\mbox{Yb}^+$ ions in a linear Paul trap. We measure a frequency fluctuation of 1.7 MHz RMS over 1000 s, and a frequency drift of 20 MHz over 7 days. Our method is suitable for use in quantum information processing experiments with trapped ion crystals.
\end{abstract}


\maketitle
\newpage

\begin{figure}[htb]
\centerline{\includegraphics*[width=8.3cm]{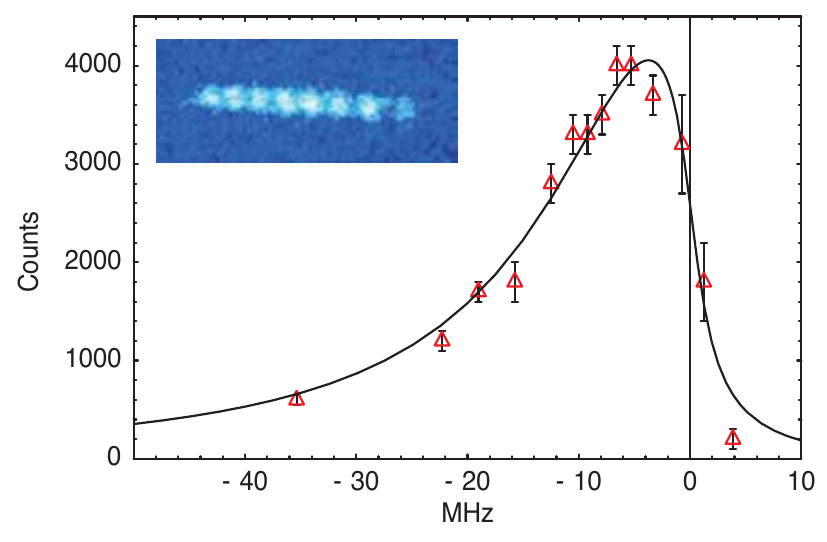} }
\caption{Fluorescence response of the ion crystal to changes in the laser frequency. 2 s accumulation time. Inset: Fluorescence image of a Coulomb crystal of trapped $^{174}\mbox{Yb}^+$ ions laser-cooled with our stabilized 369.5 nm system.}
\label{FigIonCrystalAndResponse}
\end{figure}

Experiments in atomic physics often require lasers that are tuned to within a few MHz of an atomic transition and that remain frequency-stable for hours \cite{Metcalf-vanderStraten-laser-cooling-BOOK}. In the past, these experiments have generally used optical transitions in the visible or near infrared, where there are a variety of well-established spectroscopic techniques that achieve MHz-level frequency stability with atomic or molecular vapor cells \cite{Demtroder-laser-spectroscopy-BOOK} or other neutral-atom-based secondary frequency references. The availability of blue and ultraviolet (UV) diode lasers has driven interest in their application to atomic physics experiments, stimulating interest in adapting supporting technologies such as frequency stabilization. However at these wavelengths there exist relatively few secondary frequency references compared to the wealth of atomic and molecular transitions at wavelengths greater than 500 nm. Fabry-Perot cavities are often used as a substitute when an atomic secondary frequency reference is not available, however these depend upon either a separate stabilized reference laser or the challenging engineering problem of obtaining an acceptably low drift rate. As a solution to this problem we demonstrate the frequency stabilization of a UV laser to ions in a discharge lamp and perform an independent characterization of the stability against a laser cooled ion crystal in a RF Paul trap (Fig. \ref{FigIonCrystalAndResponse} inset). While our interest is in the reliable laser cooling and trapping of ions for quantum information processing, this technique can also be used improve the precision in a wide range of ion spectroscopy related measurements, e.g., velocity-resolving LIDAR of ions in the upper atmosphere \cite{Mathews-atmosphere-ion-layer-dynamics} or Stark spectroscopy of electrical discharges \cite{Lister-Godyak-discharge-lamp-rev}. Our method also provides a straightforward way to frequency stabilize ion lasers, such as $\mbox{Ar}^+$ or $\mbox{Kr}^+$ lasers, which have narrow gain spectrums that may not overlap with a convenient transition.

Laser stabilization by optical probing of neutral atoms in a discharge was first demonstrated in 1970 \cite{Skolnick-optogalvanic-laser-stabilization} and has been extended to a variety of atomic and molecular species, including neutral Yb \cite{Kim-Yoon-Yb-DAVLL-stabilized-diode}. Spectroscopy of ions in discharges has been performed since the 1960s \cite{Hannaford-private-comm,MartenssonPendrill-Hannaford-YbII-isotope-shift} and lasers have been locked to such spectroscopic signals \cite{Hirano-Yoda-SrII-laser-stabilization}. However, an independent measurement of the frequency stability of these schemes has been lacking until now. The complicated ion production mechanism, the extremely large collision cross-section of ions as compared to neutrals, and the plasma dynamics of the discharge all lead to natural skepticism that ions in discharges can be used as MHz precision optical frequency references. Our work unambiguously shows that this level of stability can be obtained for ions in a discharge lamp. In particular, our method provides adequate stability for the production of ion Coulomb crystals, which are of great interest for quantum information processing \cite{Monroe-atom-photon-QC-rev} and as atomic clocks \cite{Hollberg-Bergquist-optical-frequency-rev}. Such crystals exhibit narrow optical transitions with natural lifetime limited linewidths (20 MHz for Yb$^+$) and frequency stability of a few MHz per year, mostly driven by magnetic field changes.

We lock a single-frequency laser at 369.5 nm to $\mbox{Yb}^+$ ions generated in a hollow-cathode discharge lamp and independently characterize the laser frequency stability by observing fluorescence from a small number of laser-cooled $\mbox{Yb}^+$ ions in a linear Paul trap. We use a form of Zeeman polarization spectroscopy \cite{Cheron-Sorel-Zeeman-effect-laser-locking}, commonly referred to as dichroic atomic vapor laser locking (DAVLL) \cite{Corwin-Wieman-DAVLL}, to derive a frequency-locking signal from the discharge. This technique is widely used for laser stabilization to neutral-atom vapor cells and has been shown to provide MHz-level long-term frequency stability \cite{Corwin-Wieman-DAVLL,Reeves-Sackett-DAVLL-temperature-stability}. Surprisingly, we find that our frequency reference has a stability similar to that of the neutral-atom systems, despite the much harsher and more complex environment experienced by ions in discharges.

The 369.5 nm laser system is an improved version of that described previously \cite{Kielpinski-Kaertner-diode-YbII-cooling}, and consists of an external-cavity diode laser prestabilized to a Fabry-Perot cavity. A UV laser diode (Nichia NDHU110APAE3) with a room-temperature operating wavelength of 371.4 nm is temperature-tuned to 369.5 nm by cooling to $-18 \:^\circ\mbox{C}$. The external cavity consists of a 3600 groove/mm holographic grating in Littrow configuration with 50\% diffraction efficiency and has a free spectral range (FSR) of 7 GHz. Rotation of the grating provides wavelength tuning over $\pm$0.9 nm. We obtain approximately 4 mW output under normal operating conditions. The grating is mounted on a piezo (Noliac CMAP-4) for fine frequency tuning; synchronously modulating the laser diode current with the piezo voltage yields a mode-hop free tuning range of 33 GHz. In comparison with the previous configuration \cite{Kielpinski-Kaertner-diode-YbII-cooling}, the shorter free-running wavelength of the laser diode at room temperature increases the operating temperature by 5 $^\circ$C and removes the need for an antireflection coating on the diode facet. We prestabilize the 369.5 nm laser frequency on short timescales by locking to a near-confocal Fabry-Perot (FP) cavity of 1 GHz FSR. An optically balanced sidelock with bandwidth 7.5 kHz reduces the laser linewidth to 1 MHz (100 ms integration time). The FP-locked laser has a residual drift of $\sim 0.1$ MHz $\mbox{s}^{-1}$ and a scan range of 11 GHz (limited by the FP piezo).

\begin{figure}[htb]
\centerline{\includegraphics*[width=8.3cm]{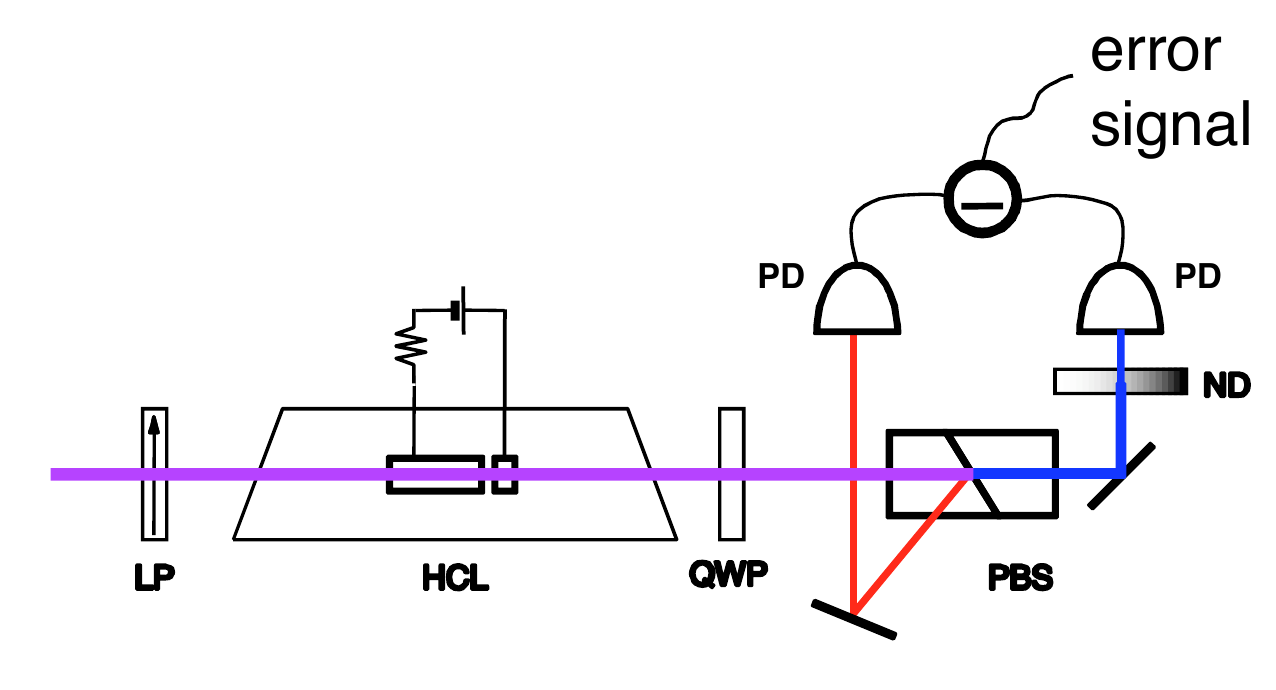} }
\caption{Atomic spectrometer optical layout. LP, linear polarizer; HCL, hollow cathode lamp; QWP, zero-order quarter waveplate; PBS, Calcite Glan-Thompson polarizing beamsplitter; ND, variable neutral-density filter; PD, photodiode.}
\label{layout}
\end{figure}

The optical arrangement of the atomic spectrometer is shown in Fig.~\ref{layout}. $\mbox{Yb}^+$ ions are produced in a DC hollow-cathode discharge lamp backfilled with Ne buffer gas (Hamamatsu H2783-70NE-Yb, 5 Torr Ne) operating at its maximum rated current (10 mA). We observe an absorption line at 369.525 nm of width $\sim 3$ GHz and maximum absorption of 6\%. The absorption line is Doppler and pressure broadened, and contains overlapping lines from the multiple Yb isotopes. Each isotope peak has a measured FWHM (full width half maximum) of 1.2 GHz, with Doppler and pressure broadening estimated to contribute approximately equal amounts. Measurement of the saturation intensity in the lamp showed no deviations from linearity to within 5\% for intensities up to 9 $I_{\mbox{sat}}$ (natural linewidth saturation intensity), indicating a pressure broadened width $>$ 700 MHz. This is consistent with the absence of saturated absorption peaks in pump/probe spectroscopy measurements of the lamp. The observed 1.2 GHz FWHM for each isotope can be found from a Voigt profile that combines a 700 MHz Lorentzian pressure broadened width with a 760 MHz Gaussian broadening width, which happens to correspond to the room temperature Doppler width. The ion density depends critically on the buffer gas composition \cite{MartenssonPendrill-Hannaford-YbII-isotope-shift}; in a He-filled lamp of identical make, we do not detect any $\mbox{Yb}^+$ absorption at all. We mount permanent magnets to each end of the lamp to produce a nominal 90 mT field parallel to the laser direction. However, our measurements indicate a Zeeman effect of the strength expected at 20 mT, a reduction similar to previously reported results \cite{Kim-Yoon-Yb-DAVLL-stabilized-diode} which may be due to magnetization of the discharge plasma or of the hollow cathode material.

On its way to the spectrometer, the 369.5 nm beam used for stabilization (the ``probe beam'') is shifted +170 MHz with a double-pass through a fused silica acousto-optic modulator (AOM). Switching the AOM drive power at 100 kHz chops the probe beam and we recover the error signal from the balanced detector (Fig.~\ref{layout}) with a lock-in amplifier (time constant 30 ms). The recovered error signal is insensitive to changes in the background photocurrents of the detectors. When the two detectors receive nearly equal photocurrents from the probe beam, i.e., near a zero-crossing of the error signal, the error signal is nearly independent of common-mode changes in the probe beam power. We balance the photocurrents to within 1\% to ensure this independence, using a variable neutral-density filter (ND, Fig.~\ref{layout}) on a micrometer-driven rotation stage. We shift the zero crossing point of the error signal to the ion resonance frequency by changing the rotation of the quarter wave plate (QWP) while maintaining the balance of photocurrents with the ND filter \cite{Yashchuk-Davis-dc-zeeman-lock,Reeves-Sackett-DAVLL-temperature-stability}. This procedure guarantees that our zero-crossing frequency is first-order independent of ion density \cite{Reeves-Sackett-DAVLL-temperature-stability} and optical power, providing additional resilience against fluctuations and gives us a tuning range of $>500$ MHz for the zero-crossing frequency. This compensation is sufficiently robust that a 40\% change in the absorption shifts the lock point by only 10 MHz and a 10\% change in the optical power shifts it by 8 MHz. For our QWP setting, we observe the trapped $^{174}\mbox{Yb}^+$ ion resonance by electronically offsetting the DAVLL signal by $<5$\% of the peak-to-peak signal. Offsets at this level maintain sufficient insensitivity to laser power and plasma density that the laser frequency stability is unaffected. We calibrate the error signal relative to the frequency markers provided by the 369.5 nm Fabry-Perot, obtaining a error slope of 1.53(3) V $\mbox{GHz}^{-1}$ for a spectrometer input power of $340 \:\mu\mbox{W}$. When locked the DAVLL signal has an in-loop noise of 0.4 mV RMS (1 s), equivalent to a frequency stability of 250 kHz and a signal to noise ratio of 2000.

\begin{figure}[htb]
\centerline{\includegraphics*[width=8.3cm]{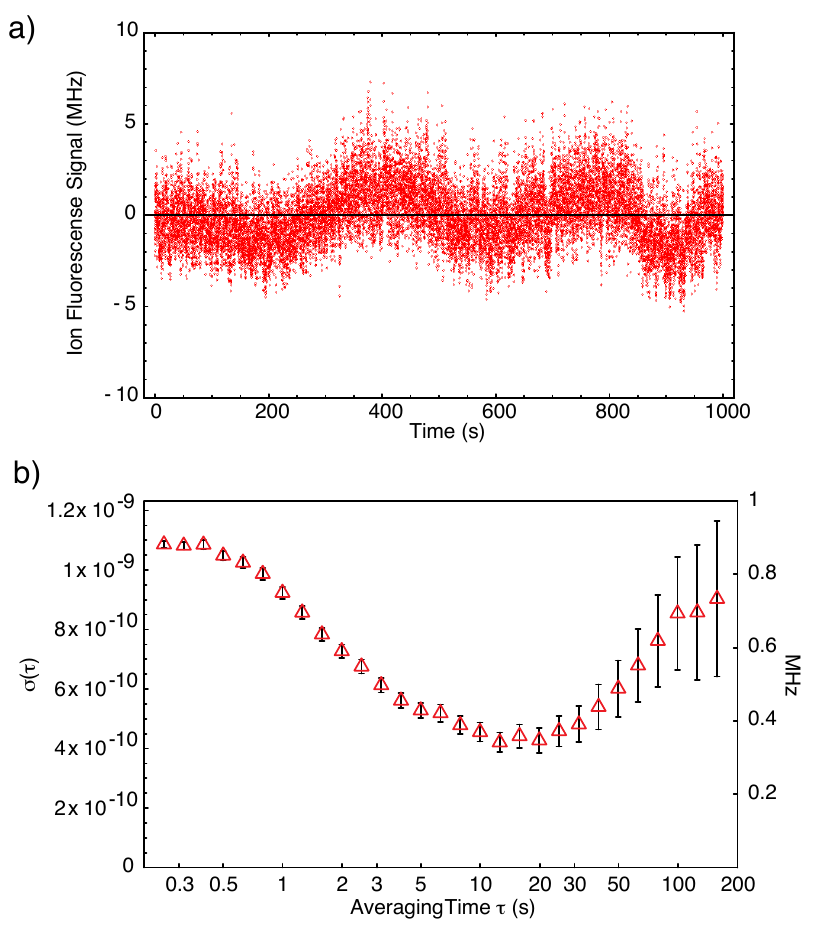} }
\caption{a) Frequency variation in the stabilized laser over 1000 s determined from variations in the ion fluorescence signal, normalized with respect to the incident laser power. b) Fractional frequency fluctuation (Allan variance  $\sigma(\tau)$) of the stabilized laser as a function of averaging time $\tau$. Results are for the combination of two 1000 s runs.}
\label{FigLaserStability}
\end{figure}

We use the fluorescence response of laser-cooled trapped $^{174}\mbox{Yb}^+$ ions as a frequency discriminator to characterize the stability of the locked 369.5 nm laser. The ion trap is identical to that used in \cite{Kielpinski-Kaertner-diode-YbII-cooling}, except that the drive frequency is 2.6 MHz, the RF voltage is 2.5 kV peak-to-peak, and the endcap voltage is 17 V. We load the trap by isotope-selective photo-ionisation using a 399 nm external-cavity diode laser resonant with a neutral Yb transition, after the manner of \cite{Balzer-Wunderlich-YbII-trap}. The 369.5 nm transition between the $^2\mbox{S}_{1/2}$ and $^2\mbox{P}_{1/2}$ states is not quite closed, with 1 out of every 200 decays \cite{Yu-Maleki-YbII-D-lifetimes} terminating in the metastable $^2\mbox{D}_{3/2}$ state. We repump the ions back to the $^2$S$_{1/2}$ state by saturating the $^2\mbox{D}_{3/2}$ - $^3\mbox{D}[3/2]_{1/2}$ transition \cite{Bell-Klein-YbII-935-repump} with an external-cavity diode laser at 935.2 nm. The photon cross-correlation technique of \cite{Berkeland-Wineland-photon-correlation-umotion} was used for minimizing the micromotion of the trapped ions. We measure lineshapes on the cooling transition (Figure \ref{FigIonCrystalAndResponse}) by adjusting the error signal offset and counting ion fluorescence over 2 s intervals. The lineshapes are typical of trapped ions laser-cooled to mK temperatures: a Lorentzian profile on the red side of resonance, with fluorescence sharply dropping off to zero as the laser frequency crosses to the blue side of resonance. Three such lineshape measurements at $\sim 1000$ s intervals yield Lorentzian linewidths of 28(3), 25(4), and 28(3) MHz, reasonably close to the natural linewidth of 19.9(5) MHz \cite{Lowe-Hannaford-YbII-lifetimes}. With the laser stabilized to the half maximum fluorescence point on the red detuned side of the resonance, the ion fluorescence response is linear in frequency to within 5\% for frequency fluctuations of less than $\pm7$ MHz.

Fig.~\ref{FigLaserStability}a shows the frequency variation of the stabilized 369.5 nm laser from the ion fluorescence over a 1000 s interval. This is calculated by dividing the ion fluorescence signal by the laser power delivered to the ions and scaling into the frequency basis according to the calibration from Fig. \ref{FigIonCrystalAndResponse}. Changes in the laser power delivered to the ions arise from stress birefringence in a fiber optic cable interacting with the polarization optics used for variable attenuation, amounting to up to a factor of 2 variation over 1000 s. Nevertheless, as expected below saturation intensity, the normalized fluorescence signal is unaffected by the drift in laser power. When the laser is unlocked near resonance, we typically observe a complete loss of ion fluorescence within 100 s.  The data of Fig~\ref{FigLaserStability}a shows a 1.7 MHz RMS fluctuation in the laser frequency over 1000 s. To obtain a more informative measure of the frequency stability, we calculate the Allan variance at averaging time $\tau$ (Fig.~\ref{FigLaserStability}b) from two sets of time-series data. The minimum Allan variance, for $\tau = 10$-20 s, corresponds to a 350 kHz effective linewidth. As an additional check, we note that the fits to our three ion lineshape measurements indicate a variation of 0.7(4) MHz in resonance frequency over a 2000 s interval. Although our infrequent changes to the setup make a strict long-term drift measurement difficult, we have observed day to day changes in the lock point of $<$10 MHz and a drift of $<$20 MHz over 7 days in a laboratory environment with better than 1$^o$C temperature  stability. Operating the Yb lamp at its maximum rated current gave us a useful operating lifetime of 30,000 mA hours  (catalog specification $>$5000 mA hours), over which the optical transmission declined from 70\% to 30\%.

In conclusion, we have demonstrated  frequency stabilization of a laser to the optical signal from ions in a discharge, and have given the first independent measurement of the laser frequency stability in such a scheme. The technique provides a tuning range of $>500$ MHz, frequency stability 350 kHz ($4.5 \times 10^{-10}$ at 369.5 nm) over 20 s, and RMS frequency fluctuations of 1.7 MHz over a 1000 s interval. The frequency drifts 20 MHz over 7 days. This technique should be readily applicable to ions of other atomic and molecular species, greatly increasing the number of transitions available for frequency stabilization, particularly in the UV band, which suffers from a shortage of secondary atomic frequency references. The technique should prove useful in atomic physics and in many other spectroscopic applications, ranging from studies of the upper atmosphere to investigations of plasma dynamics.

This work was supported by the US Air Force Office of Scientific Research under AOARD contract FA4869-06-1-0045, by the Australian Research Council (ARC) under grant DP0773354, and by H. Wiseman's ARC grant FF0458313. We thank Thorlabs Inc. for providing 370 nm AR-coated aspheric lenses. We acknowledge helpful discussions with P. Hannaford and M. Cetina.

\end{document}